\documentstyle[epsf,graphics,a4,12pt]{article}
\begin{document}
\newcommand{\beq}{\begin{equation}}
\newcommand{\eeq}{\end{equation}}
\newcommand{\beqn}{\begin{eqnarray}}
\newcommand{\eeqn}{\end{eqnarray}}
\newcommand{\dpf}{\displaystyle\frac}
\newcommand{\no}{\nonumber}
\newcommand{\ep}{\epsilon}

\begin{center}
{\Large Support of dS/CFT correspondence from perturbations of three
dimensional spacetime} 
\end{center}\vspace{1ex}
\centerline{\large 
\ Elcio Abdalla $^{a,}$\footnote[1]{e-mail:eabdalla@fma.if.usp.br},
Bin Wang$^{b,}$\footnote[2]{e-mail:binwang@fudan.ac.cn},
A. Lima-Santos $^{c}$ and W. G. Qiu $^{b}$  
}
\begin{center}
{$^{a}$ Instituto de Fisica, Univ. Sao Paulo,
C.P.66.318, CEP
05315-970, Sao Paulo, Brazil \\
$^{b}$ Department of Physics, Fudan University, Shanghai 200433,
P. R. China \\
$^{c}$ Dept. de Fisica, Univ. Fed. Sao Carlos, CP 676, CEP 13569-905, Sao
Carlos, Brazil 
}
\end{center}

\vspace{6ex}
\begin{abstract}
We discuss the relation between bulk de Sitter three-dimensional spacetime
and the corresponding conformal field theory at the boundary, in the
framework of the exact quasinormal mode spectrum. We  show that the
quasinormal mode spectrum corresponds exactly to the spectrum of thermal
excitations of Conformal Field Theory at the past boundary $I^-$, together
with the spectrum of a Conformal Field Theory at the future boundary
$I^+$.  
\end{abstract}
\vspace{6ex} \hspace*{0mm} PACS number(s): 04.06.Kz, 04.70.Dy
\vfill
\newpage

The study of quasinormal modes has been an intriguing research activity
for the last 30 years \cite{kok}, leading to important contributions to
the understanding of black holes [2-11]. 
Recently the case of Anti-de
Sitter (AdS) space has been particularly focused due to its proposed
relation to the conformal field theory (CFT) \cite{Maldacena}. Qualitative
correspondences between quasinormal modes in AdS spaces and the decay of
perturbations in the due CFT have been obtained [5-10]. More encouraging,
in the three-dimensional (3D) BTZ black hole model \cite{BTZ}, a precise
quantitative agreement between the quasinormal frequencies and the
location of poles of the retarded correlation function of the
corresponding perturbations in the dual CFT has been presented
\cite{Bir}. This gives a further evidence of the correspondence between
gravity in AdS spacetime and quantum field theory at the boundary. 

There has been an increasing interest in gravity on de Sitter (dS)
spacetimes in view of recent observational support for a positive
cosmological constant. A holographic duality relating quantum gravity on
D-dimensional dS space to  
CFT on (D-1)-sphere has been proposed \cite{Stro}. It is of interest to
extend the study in \cite{Bir} to dS space by displaying the exact
solution of the quasinormal mode problem in the dS bulk space and
exploring its relation to the CFT theory at the boundary. This could serve
as a quantitative test of the dS/CFT correspondence. This is the
motivation of the present paper. We will concentrate on nontrivial 3D dS
spacetimes. The mathematical simplicity in these models renders all
computations analytical. As shown in \cite{Carlip} \cite{MS} 3D gravity is
directly related to the two-dimensional (2D) WZW theory at the border
\cite{abdalla}, where the whole theoretical apparatus of CFT can be fully
used to obtain exact results. In such a case, Brown and Henneaux
\cite{BH} already obtained one relation between 3D AdS gravity and
conformal algebra with a (classical version of the) central charge,
recently developed into the quantum theory by Maldacena \cite{Maldacena}.

The metric of the 3D rotating dS spacetime is given by\cite{park}
\beq   \label{eq(1)}
ds^{2}=-(M-\dpf{r^2}{l^2}+\dpf{J^2}{4r^2})dt^2+(M-\dpf{r^2}{l^2}+
\dpf{J^2}{4r^2})^{-1}dr^2+r^2(d\varphi-\dpf{J}{2r^2}dt)^2,
\eeq
where $J$ is associated to the angular momentum. The horizon of such
spacetime can be obtained from 
\beq   \label{eq(2)}
M-\dpf{r^2}{l^2}+\dpf{J^2}{4r^2}=0.
\eeq
The solution is given in terms of $r_+$ and $-ir_-$, where $r_+$
corresponds to the cosmological horizon and $-ir_-$ here being imaginary,
has no physical interpretation in terms of a horizon. Using $r_+$ and
$r_-$, the mass and angular momentum of spacetime can be expressed as 
\beq  \label{eq(3)}
M=\dpf{r_+^2-r_-^2}{l^2}, \hspace{1cm} J=\dpf{-2r_+r_-}{l}
\eeq

We are now considering the problem of scalar perturbations of such a
space-time. This is what we mean here by quasinormal modes by analogy with
the analogous case of black hole perturbations. Although here we do not
have black holes the perturbations play a similar role, if we understand
the problem as perturbations of a given space-time solution. Perturbations
are not only pertinent to black holes, but can appear in any cosmological 
solution.


Scalar perturbations of this spacetime are described by the wave equation
\beq \label{eq(4)}
\dpf{1}{\sqrt{-g}}\partial _{\mu }( \sqrt{-g}g^{\mu \nu }\partial
_{\nu }\Phi) -\mu^2\Phi =0,  
\eeq
where $\mu $ is the mass of the field. Adopting the separation 
\beq  \label{eq(5)}
\Phi (t,r,\varphi )=R(r)\ e^{-i\omega t}\ e^{im\varphi },
\eeq
the radial part of the wave equation can be written as 
\beq    \label{eq(6)}
\dpf{1}{g_{rr}}\dpf{d}{rdr}( \dpf{r^{2}}{g_{rr}}\dpf{dR}{rdr}) 
+[ \omega ^{2}-\dpf{1}{r^{2}}m^{2}(
M-\dpf{r^{2}}{l^{2}})-\dpf{J}{r^{2}}m\omega] R=\dpf{1}{g_{rr}}\mu^2 R, 
\eeq
where $g_{rr}=(M-r^2/l^2+J^2/(4r^2))^{-1}$. Employing (\ref{eq(3)}) and 
defining $z=\dpf{r^2-r_+^2}{r^2-(-ir_-)^2}$, the radial wave equation can
be simplified into 
\beq    \label{eq(7)}
(1-z)\dpf{d}{dz}( z\dpf{dR}{dz}) +\left[ \dpf{1}{z}\left( 
\frac{\omega l^{2}r_{+}+mlr_{-}}{2(r_{+}^{2}+r_{-}^{2})}\right) ^{2}-\left( 
\frac{-\omega l^{2}ir_{-}+imlr_{+}}{2(r_{+}^{2}+r_{-}^{2})}
\right) ^{2}+\frac{1}{4(1-z)}\mu^2l^2 \right] R=0.
\eeq
We now set the Ansatz
\beq    \label{eq(8)}
R(z)=z^{\alpha }(1-z)^{\beta }F(z), 
\eeq
and Eq. (\ref{eq(7)}) can be transformed into
\beqn   \label{eq(9)}
z(1-z)\frac{d^{2}F}{dz^{2}} & + & \left[ 1+2\alpha -(1+2\alpha +2\beta )z%
\right] \frac{dF}{dz}  \no \\
& + & [(\beta (\beta -1)+\frac{\mu^2 l^2
}{4})\frac{1}{1-z}+\frac{1}{z}\left[ \left( \frac{\omega
l^{2}r_{+}+mlr_{-}}{2(r_{+}^{2}+r_{-}^{2})}\right) ^{2}+\alpha ^{2}\right]
\no \\ 
& - & \left[ \left( \frac{-i\omega l^{2}r_{-}+imlr_{+}}{%
2(r_{+}^{2}+r_{-}^{2})}\right) ^{2}+\alpha ^{2}+(1+2\alpha )\beta +\beta
(\beta -1)\right] ]F =0.
\eeqn
Comparing with the standard hypergeometric equation
\beq   \label{eq(10)}
z(1-z)\frac{d^{2}F}{dz^{2}}+[c-(1+a+b)z]\frac{dF}{dz}-abF=0,
\eeq
we have
\beqn    \label{eq(11)}
c & = & 1+2\alpha   \no \\
a+b & = & 2\alpha +2\beta  \no \\
\alpha ^{2}+\left( \frac{\omega l^{2}r_{+}+mlr_{-}}{2(r_{+}^{2}+r_{-}^{2})}%
\right) ^{2} & = & 0 \\
\beta (\beta -1)+\frac{\mu^2l^2 }{4} &= &0  \no \\
ab & = &\left( \frac{-\omega
l^{2}ir_{-}+imlr_{+}}{2(r_{+}^{2}+r_{-}^{2})}\right) ^{2}+(\alpha +\beta
)^{2}. \no 
\eeqn
Without loss of generality, we can take
\beq       \label{eq(12)}
\alpha =-i\left( \frac{\omega l^{2}r_{+}+mlr_{-}}{2(r_{+}^{2}+r_{-}^{2})}%
\right) ,\qquad \beta =\frac{1}{2}\left( 1-\sqrt{1-\mu^2 l^2 }\right), 
\eeq
which leads to
\beqn     \label{eq(13)}
a & = & -\frac{i}{2}\left( \frac{\omega
l^{2}+iml}{r_{+}+ir_{-}}+i(1-\sqrt{1-\mu^2 l^2  })\right),  \no \\
b & = & -\frac{i}{2}\left( \frac{\omega
l^{2}-iml}{r_{+}-ir_{-}}+i(1-\sqrt{1-\mu^2 l^2  })\right), \no \\
c & = & 1-i\left( \frac{\omega l^{2}r_{+}+mlr_{-}}{r_{+}^{2}+r_{-}^{2}}\right),
\eeqn
and the solution of (\ref{eq(7)}) reads
\beq  \label{eq(14)}
R(z)=z^{\alpha }(1-z)^{\beta }\ _{2}F_{_{1}}(a,b,c,z)
\eeq
Using basic properties of the hypergeometric equation we write the result as 
\beqn  \label{eq(15)}
R(z) &=&z^{\alpha }(1-z)^{\beta }(1-z)^{c-a-b}\frac{\Gamma (c)\Gamma (a+b-c)%
}{\Gamma (a)\Gamma (b)}\ _{2}F_{_{1}}(c-a,c-b,c-a-b+1,1-z)  \no \\
& &+z^{\alpha }(1-z)^{\beta }\frac{\Gamma (c)\Gamma (c-a-b)}{\Gamma
(c-a)\Gamma (c-b)}\ _{2}F_{_{1}}(a,b,a+b-c+1,1-z)
\eeqn
The first term vanishes at $z=1$, while the second  vanishes provided that
\beq       \label{eq(16)}
c-a=-n,\qquad or\qquad c-b=-n,
\eeq
where $n=0,1,2,..$. Employing Eqs (\ref{eq(13)}), it is easy to see that the
quasinormal frequencies are 
\beqn  \label{eq(17)}
\omega _{R} &=&i\frac{m}{l}-2i\left( \frac{r_{+}-ir_{-}}{l^{2}}\right) (n+%
\frac{1}{2}+\frac{1}{2}\sqrt{1-\mu^2 l^2 })  \no \\
\omega _{L} &=&-i\frac{m}{l}-2i\left( \frac{r_{+}+ir_{-}}{l^{2}}\right) (n+%
\frac{1}{2}+\frac{1}{2}\sqrt{1-\mu^2 l^2 }).
\eeqn
Taking other values of $\alpha$ and $\beta$ satisfying (\ref{eq(11)}), 
we have also the frequencies  
\beqn
\omega _{R} &=&i\frac{m}{l}+2i\left( \frac{r_{+}-ir_{-}}{l^{2}}\right) (n+%
\frac{1}{2}+\frac{1}{2}\sqrt{1-\mu^2 l^2 })  \no \\
\omega _{L} &=&-i\frac{m}{l}+2i\left( \frac{r_{+}+ir_{-}}{l^{2}}\right) (n+%
\frac{1}{2}+\frac{1}{2}\sqrt{1-\mu^2 l^2 }),
\eeqn
\beqn
\omega _{R} &=&i\frac{m}{l}-2i\left( \frac{r_{+}-ir_{-}}{l^{2}}\right) (n+%
\frac{1}{2}-\frac{1}{2}\sqrt{1-\mu^2 l^2 })  \no \\
\omega _{L} &=&-i\frac{m}{l}-2i\left( \frac{r_{+}+ir_{-}}{l^{2}}\right) (n+%
\frac{1}{2}-\frac{1}{2}\sqrt{1-\mu^2 l^2 }),
\eeqn
\beqn
\omega _{R} &=&i\frac{m}{l}+2i\left( \frac{r_{+}-ir_{-}}{l^{2}}\right) (n+%
\frac{1}{2}-\frac{1}{2}\sqrt{1-\mu^2 l^2 })  \no \\
\omega _{L} &=&-i\frac{m}{l}+2i\left( \frac{r_{+}+ir_{-}}{l^{2}}\right) (n+%
\frac{1}{2}-\frac{1}{2}\sqrt{1-\mu^2 l^2 }). \label{eq(17a)}
\eeqn


De Sitter space is especially appropriated for finding growing modes. Due
to the existence of negative parts in the effective potential for the
scalar field equation of motion (wells) true bound states can be formed
leading to growing modes (see \cite{bachelot}). In de Sitter space with 
a (too) small cosmological constant, corresponding to the physical case 
these bound states do not appear. Here however,  limitations on the size
of the cosmological constant do not occur here.

The shape of the potential in the three dimensional dS case is

\beq\label{potential}
V(r)=m^2(M-r^2/l^2)/r^2+Jm\omega/r^2+\mu^2\lbrack
M-r^2/l^2+J^2/(4r^2) \rbrack
\eeq

\begin{figure}[tbh]
\begin{center}
\leavevmode        
\begin{eqnarray}
\epsfxsize= 8truecm\rotatebox{-90}{\epsfbox{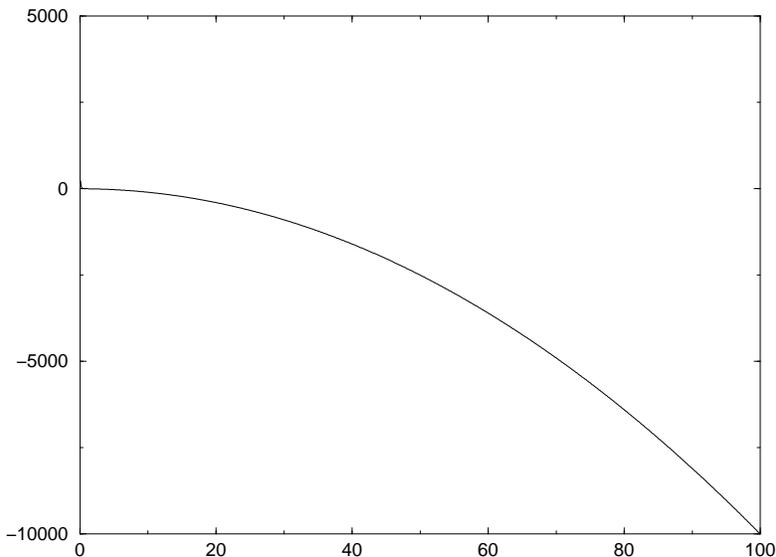}}  \nonumber
\end{eqnarray}
\vskip .5cm
\end{center}
\caption{{Potential behavior for $V(r)$, where we chose all constants 
in (\ref{potential})to be unit.  }}
\label{fig1}
\end{figure}


Let us now investigate quasinormal modes from the CFT side. It is believed
that for a thermodynamical system the relaxation process of a small
perturbation is completely determined by the poles, in the momentum
representation, of the retarded correlation function of the
perturbation. The correlation functions in dS spacetimes have been studied
in \cite{Klemm} \cite{Stro}.  

By describing the coordinates in $SO(3,1)$ \cite{Carlip} such that 
\beq \label{eq(18)}
X_1^2+X_2^2+X_3^2-T^2=l^2.
\eeq
The metric (\ref{eq(1)}) can be reobtained by the change of variables
\beqn     \label{eq(19)}
X_1 & = & l\sqrt{\chi}\sin(\dpf{r_+}{l}\varphi-\dpf{r_-}{l^2}t) \\ \no
X_2 & = & -l\sqrt{1-\chi}\cosh(\dpf{r_+}{l^2}t+\dpf{r_-}{l}\varphi) \\ 
X_3 & = & l\sqrt{\chi}\cos(\dpf{r_+}{l}\varphi-\dpf{r_-}{l^2}t)  
\label{eq(19a)} \\ \no
T   & = & -l\sqrt{1-\chi}\sinh(\dpf{r_+}{l^2}t+\dpf{r_-}{l}\varphi) \no
\eeqn
where $\chi=\dpf{r^2+r_-^2}{r_+^2+r_-^2}$.

The invariant distance between two points defined by $x$ and $x'$ reads
\cite{Stro}\cite{Klemm} 
\beq   \label{eq(20)}
d=l\arccos P,
\eeq
where $P=X^{A} \eta_{AB} X'^{B} $. In the limit $r, r' \rightarrow \infty$,
\beq \label{eq(21)}
P\approx 2\sinh \dpf{(ir_++r_-)(l\Delta\varphi-i\Delta
t)}{2l^2}\sinh\dpf{(ir_+-r_-)(l\Delta\varphi+i\Delta t)}{2l^2} \quad .
\eeq
This means that we can find the Hadamard Green's function as defined by
\cite{Klemm} in terms of $P$. Such a Green's function is defined as
$G(u,u')=<0\vert\left\{\phi(u),\phi(u')\right\}\vert 0>$ with
$(\nabla_x^2-\mu^2)G=0$. It is possible to obtain the solution 
\beq   \label{eq(22)}
G\sim F(h_+, h_-, 3/2, (1+P)/2)
\eeq
in the limit $r, r'\rightarrow\infty$, where $h_{\pm}=1\pm\sqrt{1-\mu^2 l^2}$.

Following \cite{Stro} \cite{Klemm}, we choose boundary conditions for the
fields such that  
\beq   
\lim_{r\rightarrow\infty}\phi(r,t,\varphi)
\rightarrow r^{-h_-}\phi_-(t,\varphi)\quad .
\eeq
Then, for large $r,r'$, an expression for the two point function of a
given operator $O$ coupling to $\phi$ has the form 
\beqn            \label{eq(23)}
\lim_{r\rightarrow\infty} & & \int dt d\varphi dt'
d\varphi'\dpf{(rr')^2}{l^2}\phi \stackrel{\leftrightarrow}{\partial_{r*}}
G  \stackrel{\leftrightarrow}{\partial_{r*}} \phi \\ \no 
& = & \int dt d\varphi dt' d\varphi'
\phi\dpf{1}{[2\sinh\dpf{(ir_++r_-)(l\Delta\varphi-i\Delta
t)}{2l^2}\sinh\dpf{(ir_+-r_-)(l\Delta\varphi+i\Delta t)}{2l^2}]^{h_+}}\phi
\eeqn
where $r*$ in (\ref{eq(23)}) is the tortoise coordinate.

For quasinormal modes, we have 
\beqn        \label{eq(24)}
& & \int dt d\varphi dt' d\varphi'\dpf{exp(-im'\varphi'-i\omega
't'+im\varphi+i\omega t)}{[2\sinh\dpf{(ir_++r_-)(l\Delta\varphi-i\Delta
t)}{2l^2}\sinh\dpf{(ir_+-r_-)(l\Delta\varphi+i\Delta t)}{2l^2}]^{h_+}} \no
\\ 
& \approx &
\delta_{mm'}\delta(\omega-\omega')\Gamma(h_+/2+\dpf{im/2l+\omega/2}{2\pi
T}) \Gamma(h_+/2-\dpf{im/2l+\omega/2}{2\pi T}) \no \\
& \times & \Gamma(h_+/2+\dpf{im/2l-\omega/2}{2\pi
\bar{T}})\Gamma(h_+/2-\dpf{im/2l-\omega/2}{2\pi \bar{T}}) 
\eeqn
where we changed variables to $v=l\varphi+it, \bar{v}=l\varphi-it$, and
$T=\dpf{ir_+-r_-}{2\pi l^2}, \bar{T}=\dpf{ir_++r_-}{2\pi l^2}$.  
The poles of such a correlator are 
\beqn  
\omega_L & = & -\dpf{im}{l}\pm 2\dpf{ir_+ - r_-}{l^2}(n+h_+/2), \\ \no
\omega_R & = & \dpf{im}{l} \pm 2\dpf{ir_+ +r_-}{l^2}(n+h_+/2),
\eeqn
corresponding to the quasinormal modes (\ref{eq(17)},\ref{eq(17a)})
obtained before. 

As stressed by Strominger \cite{Stro}, the boundary condition
(\ref{eq(21)}) is not mandatory, and we may choose $\phi(r,t,\varphi)\sim
r^{-h_+}\phi(t,\varphi)$ at infinity. This leads to correlations similar
to those found up to now with the change $h_+\rightarrow h_-$, and we
complete the set of quasinormal frequencies in (\ref{eq(19)},\ref{eq(19a)}). 

The dS Green functions display two symmetries, which in terms of the variable
$v$ can be described as 
\beq  \label{eq(25)}
\delta v= \dpf{2\pi l^2}{r_+^2-r_-^2}[n(r_++ir_-)+n'(r_-+ir_+)]
\eeq
for $n, n'$ arbitrary integers. We thus have a two dimensional lattice,
and a fundamental region defines the full theory. 

The quasinormal eigenfunctions thus correspond to excitation of the
corresponding CFT, being exactly those that appear in the spectrum of the
two point functions of CFT operators in dS background for large values of
$r$, that is at the boundary. 

There are some complications compared to the AdS case. There, it has well
defined temperature for the theory at the border, given by
$T^{AdS}_{\pm}=(r_+\pm r_-)/(2\pi l^2)$, where $r_+$ and $r_-$ are the
(real) values for the horizons. However, in dS case only the cosmological
horizon $r_+$ exists, while $r_-$ being imaginary. Thus the attempt to
define temperature leads us to compute values for it
$T^{dS}_{\pm}=(ir_+\pm r_-)/(2\pi l^2)$. 

It is worth noticing that since we expect instability for the black hole
formation in dS space, the Choptuik parameter $\gamma$ cannot be
defined, while in 3D AdS BTZ case it takes the value $1/2$
\cite{Birming}. We cannot apply the same method here. In particular, the
relation found in \cite{HH} between the imaginary part of the frequency,
$\gamma$ and $r_+$ in the form $\omega_{im}\sim r_+/\gamma$ cannot hold
here, since now the dS radius $l$ enters nontrivally, that is
$\omega_{im}\approx m/l + r_+(n+h_+/2)/l^2.$  We are thus in a phase where
a black hole can simply not be formed. An independent proof of this
statement would be welcome. Nevertheless, perturbations are still
pertinent, as we discussed before.

Finally, comparing with Strominger's results, we here support the picture
of CFT at a boundary of de Sitter space, finding nevertheless excitations
at both boundaries, that is, half of them in the past boundary and the
other half in the future boundary. Concluding, we found that bulk dS space
quasinormal modes can be described in terms of a thermal two-dimensional
gas with a complete set of parameters describing instabilities of the bulk
matter. The results obtained  here provide a quantitative support of the
dS/CFT correspondence. 

ACKNOWLEDGMENT:
This work was partially supported
by Fundac\~ao de Amparo \`a Pesquisa do Estado de
S\~ao Paulo (FAPESP) and Conselho Nacional de Desenvolvimento
Cient\'{\i}fico e Tecnol\'{o}gico (CNPQ). The work of B. Wang was
supported by NNSF of China.


\begin{thebibliography}{99}
\bibitem{kok} K. D. Kokkotas, B. G. Schmidt, Living Rev. Rel. 2, 2
  (1999), gr-qc/9909058.
\bibitem{price} R. H. Price, Phys. Rev. D 5, 2419 (1972); 5, 2439 (1972).
\bibitem{g-p-p} C. Gundlach, R. H. Price and J. Pullin,
  Phys. Rev. D 49, 883 (1994).
\bibitem{bick} P. R. Brady, C. M. Chambers, W. Krivan and
  P. Laguna, Phys. Rev. D 55, 7538 (1997);  P. R. Brady, C. M. Chambers,
  W. G. Laarakkers and E. Poisson, Phys. Rev. D 60, 064003 (1999).
 
\bibitem{hod} J. S. F. Chan and R. B. Mann, Phys. Rev. D 59, 064025
(1999); J. S. F. Chan and R. B. Mann, Phys. Rev. D 55, 7546 (1997).

\bibitem{HH} G. T. Horowitz and V. E. Hubeny, Phys. Rev. D 62,
024027 (2000); G. T.
Horowitz, Class. Quant. Grav. 17, 1107 (2000).

\bibitem{wang2} B. Wang, C. Y. Lin and E. Abdalla, {\it Phys.
Lett.} {\bf B 481}, 79 (2000), hep-th/0003295.
\bibitem{wang3} B. Wang, C. Molina and E. Abdalla,{\it Phys. Rev. } {\bf D 63},
084001 (2001); J. M. Zhu, B. Wang and E. Abdalla,{\it Phys. Rev. } {\bf D 63},
124004 (2001).
\bibitem{lemos} V. Cardoso, J. P. S. Lemos, {\it Phys. Rev.} 
{\bf  D 63}, 124015 (2001); gr-qc/0105103.
\bibitem{wangabdman} B. Wang, E. Abdalla and R. B. Mann, {\it Phys. Rev. 
} {\bf D 65} 084006 (2002), hep-th/0107243. 
\bibitem{Bir} D. Birmingham, I. Sachs and S. N. Solodukhin,
{\it Phys. Rev. Lett} {\bf 88} 151301 (2002), hep-th/0112055.
\bibitem{Maldacena} O. Aharony, S. S. Gubser, J. Maldacena, H. Ooguri and
Y. Oz, {\it Phys. Rept.} {\bf  323}, 183 (2000). 
\bibitem{BTZ} M. Banados, C. Teitelboim and J. Zanelli,
{\it Phys. Rev. Lett.} {\bf  69}, 1849 (1992). 
\bibitem{Stro} A. Strominger, ``The dS/CFT correspondence'', hep-th/0106113.
\bibitem{Carlip} S. Carlip, {\it Class. Quant. Grav. } {\bf 12}, 2853 (1995).
\bibitem{MS} J. Maldacena and A. Strominger, {\it JHEP} {\bf  9802} (1998) 014.
\bibitem{abdalla} E. Abdalla, MCB Abdalla and K D Rothe, ``Non
perturbative Methods in two-dimensional Quantum Field Theory'', second
edition, WS (2001). 
\bibitem{BH} J. D. Brown and M. Henneaux, {\it Commun. Math. Phys.} {\bf
104}, 207 (1986). 
\bibitem{park} M. I. Park, {\it Phys. Lett.} {\bf B 440}, 275 (1998).
\bibitem{bachelot} A. Batchelot and A. Motet-Bachelot {\it Annals
Inst. H. Poincare} {\bf 59} (1993) 3.
\bibitem{Klemm} D. Klemm, ``Some Aspects of de Sitter/CFT
Correspondence'', hep-th/0106247. 
\bibitem{Birming} D. Birmingham, {\it Phys. Rev.} {\bf  D 64}, 064024 (2001);
M. W. Choptuik, {\it Phys. Rev. Lett.} {\bf  70}, 9 (1993). 
\end{thebibliography}
\end{document}